# Particle size dependence of magnetization and phase transition near $T_N$ in multiferroic $BiFeO_3$

R. Mazumder, S. Ghosh, P. Mondal, Dipten Bhattacharya,* S. Dasgupta, N. Das, and A. Sen
*Sensor & Actuator Section, Central Glass and Ceramic Research Institute, Calcutta 700 032, India*

A.K. Tyagi
*Solid State and Surface Chemistry Section, Bhabha Atomic Research Center, Trombay, Mumbai 400 085, India*

M. Sivakumar[†]
*Ultrasonic Processing Group, National Institute of Advanced Industrial Science and Technology, Nagoya 463-8560, Japan*

T. Takami and H. Ikuta
*Department of Crystalline Materials Science, Nagoya University, Nagoya 464-8603, Japan*

We report results of a comprehensive study of the phase transition at $T_N$ (~643 K) as a function of particle size in multiferroic $BiFeO_3$ system. We employed electrical, thermal, and temperature dependent X-ray diffraction (XRD) studies in order to characterize the transition in a host of samples. We also carried out detailed magnetic measurements over a temperature regime 2-300 K under a magnetic field 100-10000 Oe both on bulk and nano-crystalline systems. While in the bulk system a sharp endothermic peak at $T_N$ together with a broad feature, ranging over nearly ~150 K *(DT)*, could be observed in calorimetry, the nanoscale systems exhibit only the broad feature. The characteristic dielectric anomaly, expected at $T_N$, is found to occur both at $T_O$ and $T_N$ across $DT$ in the bulk sample. The Maxwell-Wagner component due to interfaces between heterogenous regions with different conductivities is also present. The magnetic properties, measured at lower temperature, corroborate our observations in calorimetry. The metastability increases in the nanoscale $BiFeO_3$ with divergence between zero-field cooled (ZFC) and field cooled (FC) magnetization below ~100 K and faster magnetic relaxation. Interestingly, in nanoscale $BiFeO_3$, one also observes finite coercivity at lower temperature which points out that suitable design of particle size and shape may induce ferromagnetism. The inhomogeneous distribution of Bi/Fe-ions and/or oxygen non-stoichiometry seems to be giving rise to broad features in thermal, magnetic as well as in electrical responses.

---

*Electronic address: dipten@cgcri.res.in
[†]Present address: Bharathidasan University, Tiruchirapalli 620024, Tamil Nadu, India

## I. INTRODUCTION

The multiferroics, with, at least, two of the three orders or degrees of freedom – (anti)ferromagnetic, (anti)ferroelectric, and ferroelastic – co-existing and often a coupling among them, are rare in nature as transition metal ions with active d-electrons often tend to reduce the off-center distortion necessary for ferroelectricity.[1] It has been pointed out by a recent theoretical work[2] that additional electronic mechanism (mainly $s^2$ lone pair at another site) provides the driving force for ferroelectricity. Since their first observation by Smolenskii and co-workers[3] in 1958, three different classes of multiferroic systems could be identified - $RMnO_3$ (where R = Dy, Tb, Ho, Y, Lu, etc.), $RMn_2O_5$ (where R = Nd, Sm, Dy, Tb), and $BiBO_3$-type (where B = Mn, Fe). Such systems depict magneto-electric coupling and, therefore, magnetic polarization can be achieved by electric field and vice versa. Recently, there has been a renewed interest[4] in multiferroic systems due to the observation of orders of magnitude large magneto-electric coupling and the application potential of these systems in a range of devices based on 'spintronics'– magneto-electric sensors, electrically driven magnetic data storage and recording devices, magneto-capacitive devices, nonvolatile memories etc. Direct evidence of linear coupling between electric and magnetic domains could be observed in $YMnO_3$ using non-linear optics – optical second harmonic generation.[5] However, the temperature range over which the multiferroic behavior is observed varies from system to system. For instance, $RMnO_3$ (space group $P6_3cm$), and $RMn_2O_5$ (space group Pbam) systems depict multiferroic properties at lower temperature (<100 K).[6] In $BiMnO_3$ (space group C2) too, ferromagnetic and ferroelectric order co-exist below ~105 K. The $BiFeO_3$ system (space group R3c) is interesting, in this respect, as it depicts ferroelectric order below $T_C$ ~ 1110 K and antiferromagnetic order below $T_N$ ~ 643 K. Therefore below $T_N$, one expects coupling between electric and magnetic domains.

The multiferroic nature of $BiFeO_3$ is due to stereochemical activity associated with the $6s^2$ lone pair of $Bi^{3+}$. It results in lowering of structural symmetry and hence ferroelectricity.[7] The magnetic order is little complicated with ferromagnetic coupling within a plane and antiferromagnetic coupling between two adjacent planes. Due to antisymmetric Dzyaloshinsky-Moriya (DM) exchange interaction, canted antiferromagnetism may develop. However, a spiral spin structure is superimposed on this antiferromagnetic order in which the antiferromagnetic axis is rotated through the crystal and develops an incommensurate order.[8] The magneto-electric coefficient (dE/dH) is found to vary within 3V/cm.Oe across a temperature range 77-300 K.[9] Weak magnetization, however, restricts its large scale use in several devices. One needs to explore whether large magnetization can be induced in these systems by doping and/or structural manipulation. In other words, one needs a thermodynamically stable system with both ferromagnetism and ferroelectricity existing above room temperature. $BiFeO_3$ could not be exploited for any novel application for two major problems: (i) weak magnetization as it is essentially an antiferromagnet below $T_N$ and (ii) large loss factor because of oxygen non-stoichiometry. Therefore, even though magneto-electric coupling could be observed above room temperature, unlike $CdCr_2S_4$ [Ref. 10] complete ferroelectric polarization flop cannot be achieved in $BiFeO_3$ by magnetic field sweeping. It has recently been observed[11] that in thin film $BiFeO_3$, finite magnetization of the order of ~1$\mu_B$/Fe prevails even above room temperature. The origin of such magnetization is a bit controversial.[12] It has been proposed, however, that such finite magnetization could result from increased canting.[13] A suitable doping at Bi- and/or Mn-site, on the other hand, helps in minimizing the conductivity and hence the loss factor.[14] It will be interesting to explore whether both the problems can be addressed and tackled in nanoscale $BiFeO_3$.

In this paper, we report the results of a comprehensive study carried out both on bulk and nanoscale $BiFeO_3$ for exploring the phase transition features near $T_N$ as well as the low temperature magnetic properties. We employed dc resistivity, dielectric measurements, differential scanning calorimetry (DSC), differential thermal analyses (DTA), dilatometry, temperature dependent X-ray diffraction, and detailed magnetic measurements in order to systematically record the behaviors of the bulk and nano-system below and above the phase transition point. It appears that the local inhomogeneity as well as oxygen non-stoichiometry across the sample volume has a major role to play in governing all the properties.

## II. EXPERIMENTAL DETAILS

Primarily two types of measurements were carried out on phase pure bulk and nano-particles of $BiFeO_3$ – thermal and magnetic. The electrical measurements were carried out only on bulk sintered samples. All the samples have been prepared by different solution chemistry routes. While bulk ~1 μm size particles have been prepared by conventional co-precipitation technique, finer 4-50 nm size particles have been prepared by sonochemical and solution evaporation (tartaric acid template or ferrioxalate precursor) techniques. In the sonochemical technique, powders are prepared in solution chemistry route under ultrasonic vibration. A mixed aqueous solution of $Bi(NO_3)_3$ and $Fe(NO_3)_2$ – taken in proper stoichiometric ratio – is sonicated by using Ti horn (20 kHz, 1500 W, Vibracell, USA) till the precipitation is complete. A small amount of decalin is used for proper power transfer and SDS surfactant is also used for preventing agglomeration. The precipitated powder is collected, washed by alcohol, and dried in a vacuum oven at ~40$^o$C. In the tartaric acid template technique, aqueous solutions of metal nitrates were mixed with tartaric acid. The mixed solution was boiled at 150-160$^o$C till the liquid evaporated out. The green fluffy powder was heated for 1h before calcination. The tartaric acid template helps in forming heterometallic polynuclear polymeric arrays where metal ions come in close proximity to yield phase-pure $BiFeO_3$ after *in-situ* oxidization. If oxalic acid is used as a template, it forms Bi-ferrioxalate precursor which undergoes decomposition in presence of oxidizing agent like $HNO_3$. In this case too, heteronuclear polymeric array, with metal ions, forms. Other organic acids like citric acid do not give rise to such phase pure $BiFeO_3$ as it tends to form dimeric chain instead of polynuclear polymeric chain. It appears, therefore, that the formation of heteronuclear polymeric arrangement prior to oxidization is the key to the formation of phase-pure $BiFeO_3$. The as-prepared powders were calcined at 400-700$^o$C and the bulk samples were sintered at ~840$^o$C for 10h. Both the bulk and nanoscale samples were characterized by X-ray diffraction (XRD) and scanning electron microscopy (SEM) studies. In addition, the nano-particles have been studied by transmission electron microscopy (TEM). The specific surface area of the nano-particles has also been measured. Further details of the sample preparation and the results of the characterization are available elsewhere.[15,16] The crystallite size of the nanoscale systems could be estimated from X-ray peak broadening by Debye-Scherrer equation using the full width at half maximum (FWHM) data. The particles sizes could be estimated from BET specific surface area data as well as directly from TEM photographs. A representative TEM photograph for nanoscale $BiFeO_3$ is shown in Fig. 1.

The thermal and magnetic measurements have been carried out on powder samples while electrical measurements have been carried out on sintered pellets. The differential scanning calorimetry (DSC) study has been carried out in a Perkin-Elmer Diamond DSC over a temperature range 300-800 K while the differential thermal analysis (DTA) measurements were carried out in a Shimadzu DTA-50. The heating rate was $10^{o}C$/min in all the cases. The thermal expansion has been measured in an ORTON dilatometer (Model 1600). The low-frequency (1 Hz – 1 MHz) dielectric measurements have been carried out in a LCR meter (Hioki Hi-Tester 3652-50). The electrodes have been prepared by silver coating and the coating was cured at ~$600^{o}C$ in air.

The magnetic measurements have been carried out in a SQUID magnetometer (MPMS) of Quantum Design. The zero-field cooled (ZFC) and field-cooled (FC) magnetization vs. temperature patterns have been studied across a temperature range 2-300 K while magnetic hysteresis loops over ±10000 Oe have been recorded at several temperatures. We also carried out magnetic relaxation measurements at ~5 K and ~300 K under ~100 and ~10,000 Oe for both bulk and nanoscale samples.

## III. RESULTS AND DISCUSSION

### A. Thermal studies

In Fig. 2a, we show the specific heat ($C_p$) vs. temperature patterns – extracted from differential scanning calorimetry (DSC) and differential thermal analysis (DTA) data – for both bulk and nanoscale $BiFeO_3$. The specific heat is found to be increasing as a function of particle size. A small yet sharp peak – signifying first order transition – could be noticed at ~653 K ($T_N$) in the case of bulk sample. In the inset of Fig. 2a, we show the ($C_p/T$) vs. temperature ($T$) plot of the region around the peak for the bulk sample. The change in entropy $\Delta S$ [= $\int(C_p/T).dT$] is estimated to be ~0.0612 J/g.K. In Fig. 2b, we show the thermal expansion data for the bulk sample. The phase transition feature is clearly evident near $T_N$.

Apart from the rise in specific heat with the increase in particle size, one can also notice a variation in the phase transition feature with the particle size. For finer particles, a very broad feature – spanning nearly 150 K – appears while with the increase in particle size (≥50 nm), the broadness reduces a bit and the feature near $T_N$ becomes more prominent. The absence of a sharp peak clearly points out local inhomogeneity and broad phase transition across a wide temperature range. Although, such broad feature could be observed in bulk samples as well, simultaneous presence of sharp peak highlights more ordering in bulk sample.

The nano-crystalline samples contain primarily inhomogeneous phases where lattice defects, oxygen non-stoichiometry etc. appear to be sizable. In addition, lattice strain is also high in nano-crystalline system. The combined effect has given rise to even smaller scale of homogeneity in nano-system. As a result, at least in global calorimetry, no clear endothermic peak could be observed. In the absence of a sharp peak in nano-crystalline samples, it is difficult to identify the phase transition temperature and, hence, to figure out how the transition point has shifted as a function of particle size. The absence of a sharp peak in DSC/DTA thermograms is corroborated by the fact that temperature-dependent XRD patterns (Fig. 3) do not show any clear structural transition across $T_N$.

**B. Electrical studies**

In Fig. 4, we show the real part of the dielectric permittivity vs. temperature for the bulk sample across a temperature range 300-673 K. The impedance ($Z^*$) and modulus ($M^*$) spectra at different temperatures are shown in Fig. 5. We show here the data corresponding to the frequency range 100 Hz-1MHz. The real dielectric permittivity [*e'(w,T)*] exhibits two anomalies – at $T_O$ and $T_N$ – over a temperature span ~150 K (*DT*). The temperature $T_O$ marks the onset (~550 K) and $T_N$ marks the termination of the transition process. Along with magnetic transition, $BiFeO_3$ undergoes a structural phase transition as well, at $T_N$. The structural transition is evident from the peak in DSC thermogram at $T_N$. Therefore, one expects anomaly in dielectric permittivity due to both structural transition as well as magneto-electric coupling. Two anomalies across *DT* – instead of only one at $T_N$ – result from broadening of the magnetic and structural transition due to inhomogeneity. The dielectric permittivity exhibits high value and large frequency dispersion both below and above the phase transition range *DT*. The loss factor tanδ varies over a range 0.1-10. These features of the dielectric spectra are characteristic of Maxwell-Wagner relaxation process. The Maxwell-Wagner component arises from grain boundary interfaces in these polycrystalline samples as well as from intrinsic heterogeneity which give rise to broad magnetic transition over a span of ~150 K around $T_N$. The presence of Maxwell-Wagner relaxation can be further established by the fact that *e"(w)* tends to a high value as *w* tends to zero; in the case of Debye relaxation process *e"(w)* tends to zero as *w* approaches zero. Therefore, the overall dielectric response appears to be consisting of Maxwell-Wagner component as well as intrinsic polarization component. We analyzed the impedance spectra and the modulus (*M**) spectra in order to figure out the relaxation time scale (*t*) of the intrinsic polarization component. The extrinsic interface polarization relaxation frequency scale is found to be within 100-1000 Hz, while the intrinsic polarization relaxation frequency scale turns out to be 10 kHz – 1 MHz over a temperature range 300-650 K [Fig. 5]. The intrinsic relaxation time scale (*t*) is plotted as a function of temperature (*T*) in Fig. 6. A broad anomaly quite akin to that observed in the case of *e'(w,T)* vs. *T* (Fig. 4) is observed in the case of *t* vs. *T* pattern. Below the anomaly at $T_O$, the τ vs T pattern turns out be nearly Arrhenius while above the anomaly too, the pattern appears to be Arrhenius. In the inset of Figure 6 we show the natural log of four-probe dc resistivity vs *1/T* pattern. The activation energies from both *t-T* and resistivity vs. *1/T* pattern have been calculated. While the activation energies, calculated from resistivity data, turn out to be ~0.88 and 0.19 eV at below $T_O$ and above, they are 0.46 and 0.9 eV over the same temperature ranges when estimated from relaxation time scale vs. temperature data. The difference in the values of the activation energies, calculated from resistivity and relaxation time data, signifies that the mechanism of relaxation of the electric polarization is different from polaron hopping.

Therefore, the overall dielectric response appears to be consisting of both Maxwell-Wagner and intrinsic polarization components. Spontaneous ferroelectric order is expected with no relaxation feature in the case of intrinsic polarization. Yet we observe a relaxation along with a broad transition. These observations point out that even in the bulk sample, intrinsic inhomogeneity plays a major role. The inhomogeneity results from volatility of Bi near the processing temperature 820-850$^o$C and/or oxygen non-stoichiometry inherent in the complicated crystal chemistry of $BiFeO_3$. In fact, the polarization ($P$) vs. field ($E$) loop in the case of pure $BiFeO_3$ appears to reflect the lossy nature. The inhomogeneity results in smaller ferroelectric domains with faster relaxation dynamics. It would have been interesting to study the relaxation dynamics and transition point in nanoscale system where the scale of inhomogeneity is higher and the homogeneous cluster size is small. Since sintering leads to growth of nano-grains into micron size, no electrical measurement could be done on nano-crystalline samples.

**C. Magnetic studies**

We have measured ZFC, FC magnetization vs. temperature patterns under different fields as well as hysteresis loops at several temperatures across a range 2-300 K. The magnetic relaxation pattern has also been observed at different temperatures and under different fields. In Fig. 7, we show the ZFC and FC magnetization vs. temperature patterns for both the bulk and nano-crystalline $BiFeO_3$. The bulk system exhibits antiferromagnetic order below $T_N$ ~ 643 K. Therefore, we could not study the magnetization across the transition point. The magnetization at ~2 K is found to be ~0.016$\mu_B$/Fe atom under ~100 Oe and ~1.5$\mu_B$/Fe atom under ~10,000 Oe for the bulk sample. It drops down to ~0.008$\mu_B$/Fe atom and 0.7$\mu_B$/Fe atom, respectively at ~300 K.

On the other hand, for nanoscale sample, the low temperature magnetization (at ~2 K) is found to be nearly equal to what has been observed in the case of bulk sample. But, at ~300 K, the magnetization is higher by a factor of two. For high spin (S = 5/2) $Fe^{3+}$ ions, the effective magnetization should be ~5.92$\mu_B$. Therefore, it appears that the spin configuration per Fe ion is far from parallel. There are several other important points worthy of noting in the magnetization vs. temperature data: (i) no divergence between ZFC and FC data in the case of the bulk sample; (ii) the magnetization changes with magnetic field; (iii) in the case of nanoscale sample the ZFC and FC magnetization data diverge below ~100 K; (iv) in all the cases – bulk and nano – the magnetization rises with decrease in temperature; (v) the temperature below which the magnetization starts rising is ~50 K for the bulk sample and ~10 K for the nanoscale sample. Interestingly, in the case of bulk sample an anomaly near ~65 K could be observed in the low-field ZFC data. The FC as well as high field data, of course, does not depict any such anomaly. This seems to be because of a metastable secondary antiferromagnetic phase. The origin of such a phase is still unknown and not previously reported. The nano-particles of the $BiFeO_3$ do not depict any anomaly around ~65 K.

Because of stable antiferromagnetic order within the temperature range 2-300 K, no divergence between ZFC and FC magnetization vs. temperature plots could be noticed in the case of bulk sample. The rise in magnetization below ~50 K could be due to additional magnetic moment arising out of $Bi^{3+}$-$O^{2-}$ bonds. The nanoscale sample, on the other hand, exhibits divergence between ZFC and FC magnetization data below ~100 K which could be due to the onset of spin glass state. The magnetization at any temperature, within this regime, depends on history effect – the path followed for reaching the point. The ZFC data exhibit a broad cusp-like feature below ~100 K which is typical of nanoscale antiferromagnetic system.[19]

In Fig. 8, we show the hysteresis loops for bulk and nano-particle systems. For the bulk $BiFeO_3$ sample, the hysteresis loops at several temperatures exhibit patterns typical of an antiferromagnetic system with no hysteresis and nearly zero coercivity. The nano-particle system, on the other hand, exhibits sizable hysteresis and finite coercivity below

~100 K. In fact, one can easily notice a deviation from linearity and formation of a ferromagnetic loop at a lower field. The coercivity at ~5 K is found to be ~450 Oe. *An observation of such high coercivity in nanoscale $BiFeO_3$ is quite significant.* This could be because of lattice strain-induced spin canting or ferromagnetism in nanoscale. It has been reported earlier, that lattice strain in thin film gives rise to higher magnetism (~1 $\mu_B$/Fe). The observation of higher coercivity in nanoscale may help in finding a way for improving magnetization in $BiFeO_3$ which, in turn, will lead to many applications.

In order to find out the stability of the magnetic phases – both at low temperature (~5 K) and room temperature (~300 K) – in the nano-particle vis-à-vis bulk system, we measured the magnetic relaxation pattern over a time scale of ~36000s. The relaxation patterns are shown in Fig. 9. The amount of variation in initial magnetization ($M_0$) with time ($t$) varies with temperature, applied field, and sample type - bulk or nano-particle systems. The initial magnetization decreases with time in the case of low field (~100 Oe) while it increases with time in the case of higher field (~10000 Oe). In all the cases, however, the pattern of variation is logarithmic in nature where the magnetization at any time $t$ follows $M(t,T) = M_0(T) \pm S(T) \ln(t/t_m)$ equation; $S$ is magnetic viscosity and is related to the activation energy, $\tau_m$ is a characteristic time scale. This is a typical flux-creep pattern. The extent of relaxation in the bulk system is negligible at low temperature and is very small (~0.53%) at high temperature highlighting the stability of the phase. The nano-particle system, of course, exhibits large relaxation (~0.61%) at low temperature and nearly an order of magnitude higher relaxation (~3.7%) than that of the bulk system at high temperature. The high field relaxation patterns depict a rise in magnetization. This could be due to metastability in the system. The activation energies are found to be varying over ~0.08-0.25 eV with field (100 Oe – 10000 Oe) at ~5 K and over 1.4-10.0 eV at ~300 K. In bulk sample, of course, the variation in the magnetization is very small both at ~5 and ~300 K. The activation energies for the bulk systems are found to be enormous – of the order of ~$10^4$ eV – reflecting the stability of the magnetic phases.

These results point out that a magnetic state with a higher magnetization and a finite coercivity develop in the case of nanoscale samples. Of course, the magnetic state is metastable with lower activation energy and higher relaxation rate. The metastability is possibly due to spin frustration in nanoscale.

## IV. SUMMARY

In summary, we have studied the phase transition near $T_N$ as a function of particle size in multiferroic $BiFeO_3$ system by employing calorimetry, dielectric, and temperature dependent X-ray diffraction studies. Because of inhomogeneity, the phase transition at $T_N$ is found to become broader – both in bulk and in nanoscale samples. The inhomogeneity possibly arises from volatility of Bi near the processing temperature as well as susceptibility of the lattice structure for developing oxygen non-stoichiometry. Such broader transition gives rise to two anomalies in dielectric permittivity across a range $DT$ in the case of bulk sample. The magnetization vs. temperature data depict a rise in magnetization at lower temperature for both the cases while an onset of spin glass state in the case nanoscale sample below ~100 K. The magnetization in the latter case relaxes at a faster rate and yields lower activation energy. One interesting observation is the ferromagnetism and finite coercivity at lower temperature (<100 K) in nanoscale which possibly results from enhanced lattice strain. This result points out that suitable design of $BiFeO_3$ particles – size and shape – can improve the magnetization of the system.

## ACKNOWLEDGEMENT

This work has been carried out under the network program "custom-tailored special materials" (CMM 0022) of CSIR, Govt. of India. One of the authors (R.M.) thanks CSIR for a senior research fellowship.

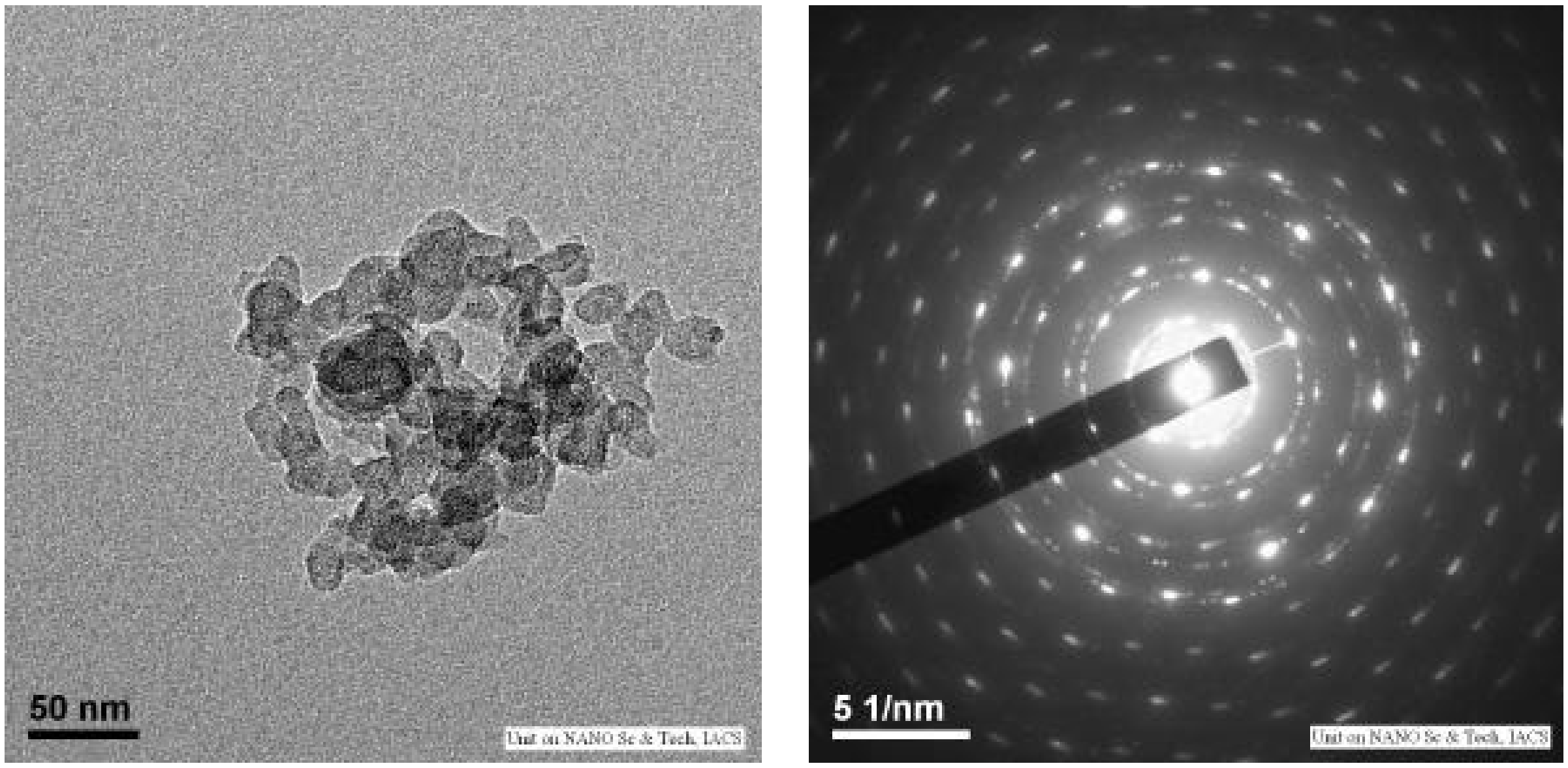

Fig. 1. Typical transmission electron microscopy (TEM) photograph of nanoscale $BiFeO_3$ system. The selected area electron diffraction (SAED) pattern is also shown. The nanoscale particles are crystalline in nature.

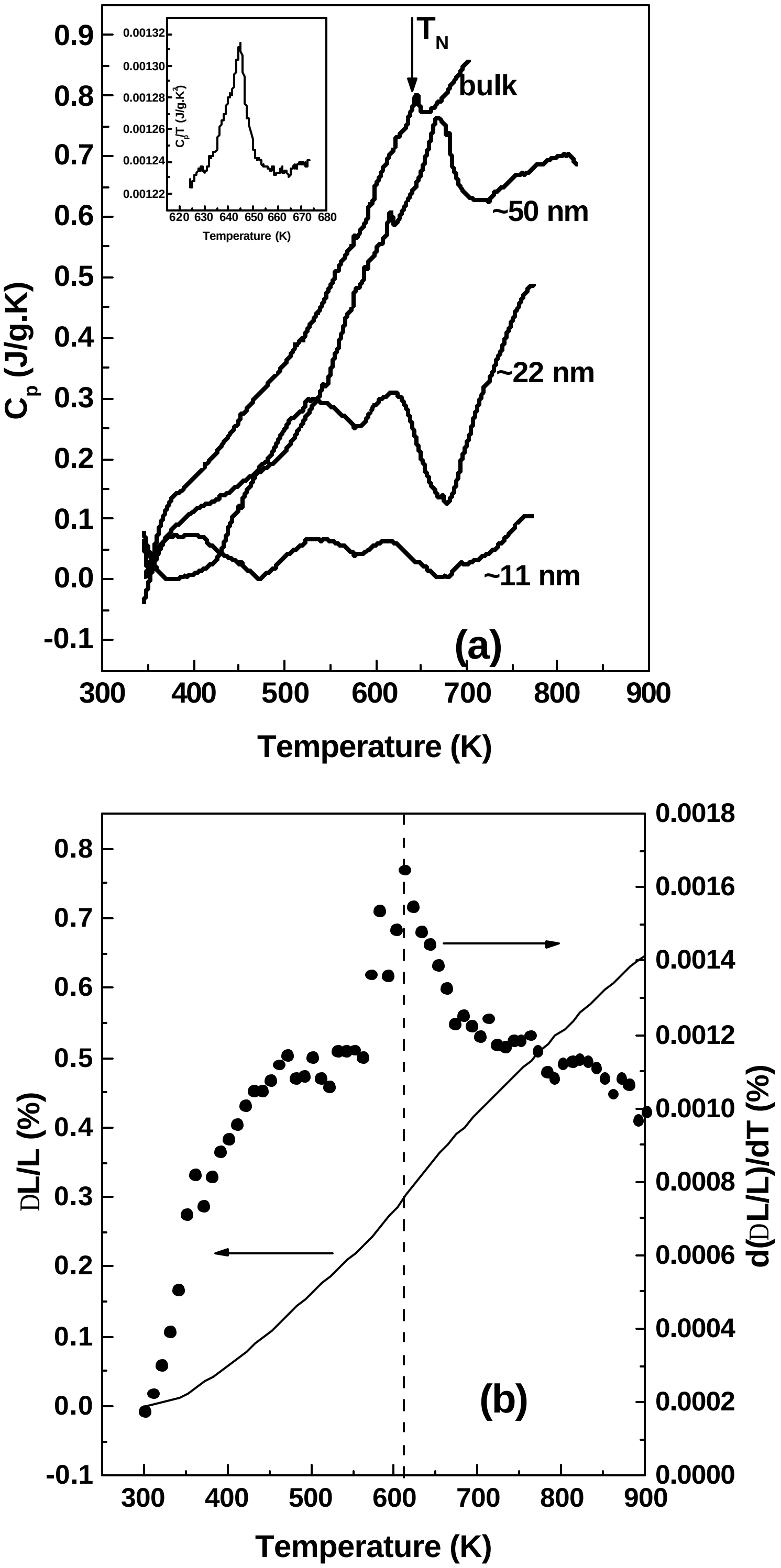

Fig. 2. (a) The specific heat vs. temperature plots for the bulk and nanoscale $BiFeO_3$; (b) the thermal expansion vs. temperature plot for the bulk $BiFeO_3$ sample. The variation of thermal expansion coefficient ($\alpha$) with temperature is also shown.

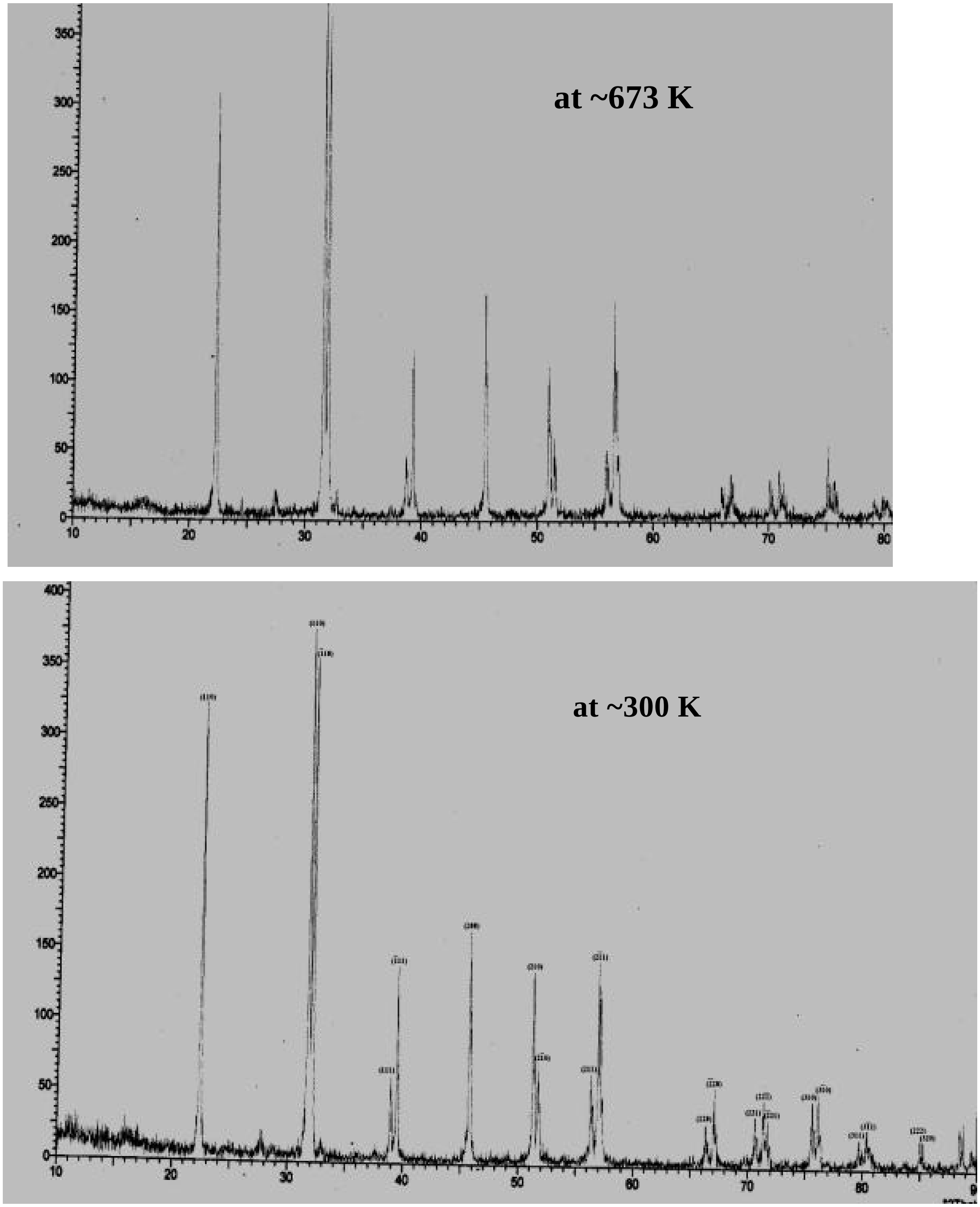

2θ

Fig. 3. X-ray diffraction pattern for nanoscale $BiFeO_3$ at room temperature and above $T_N$.

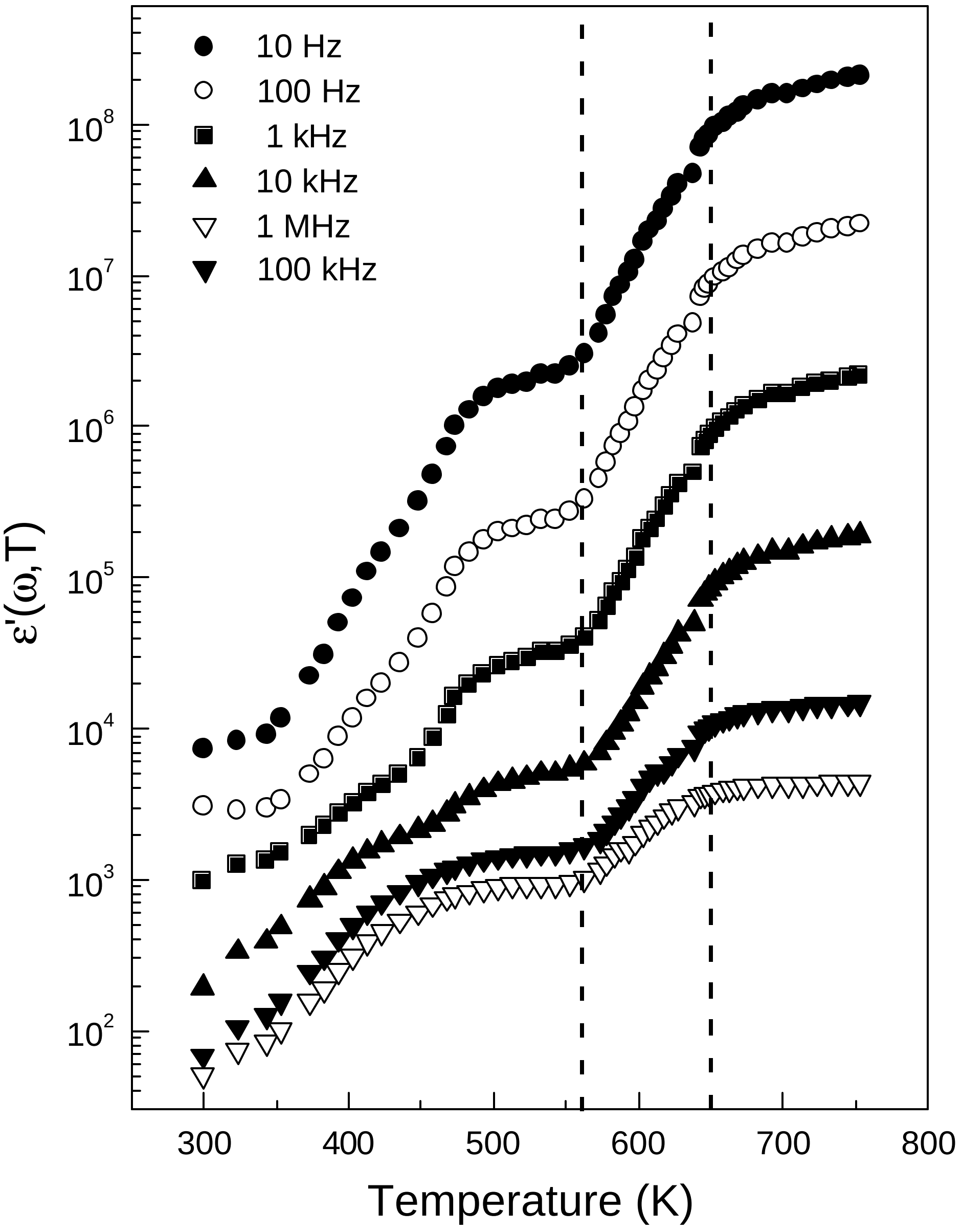

Fig. 4. Real part of the dielectric permittivity [ε'(ω,T)] vs. temperature patterns for the bulk $BiFeO_3$ system. Anomalies at $T_O$ and $T_N$ are quite evident.

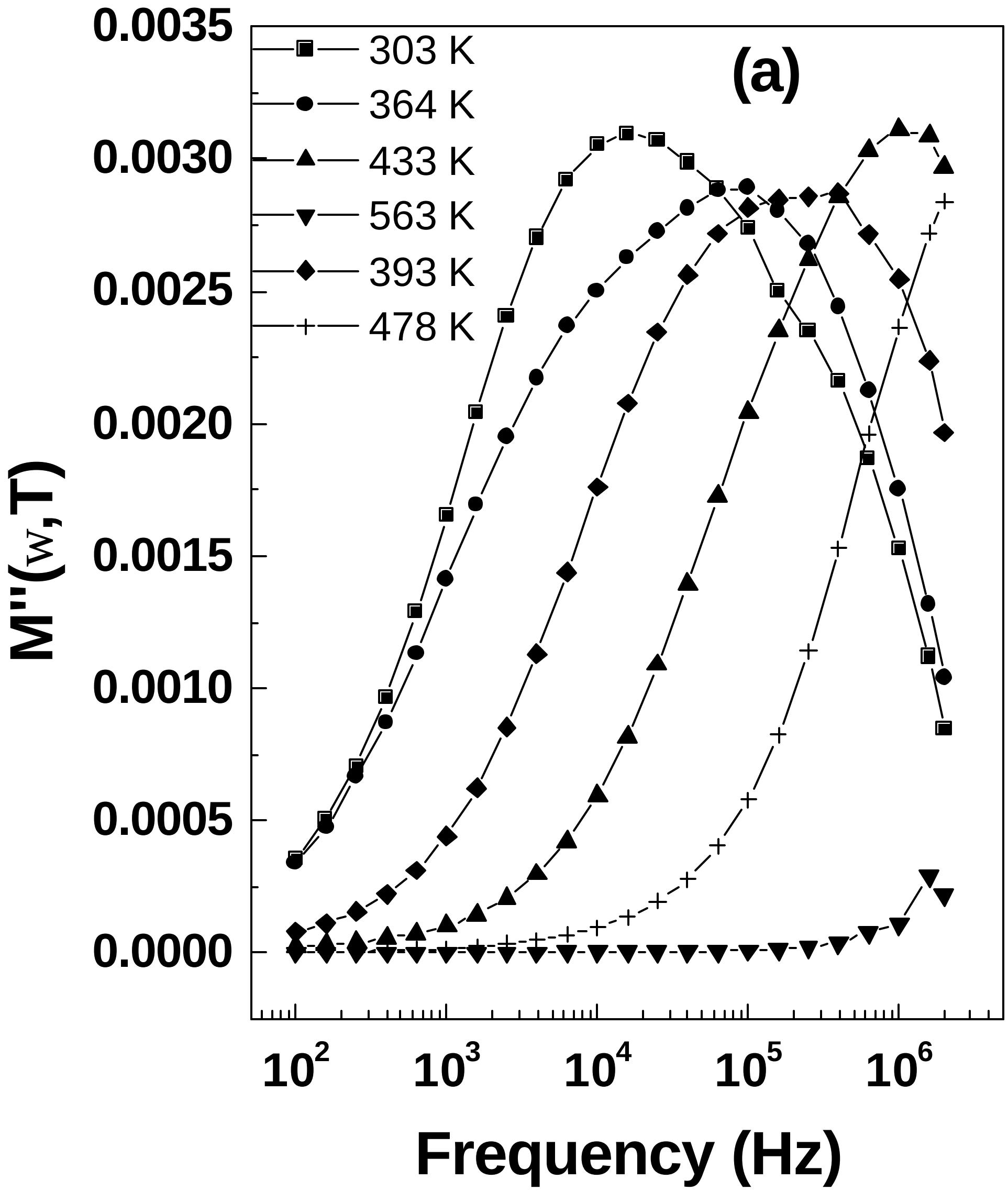

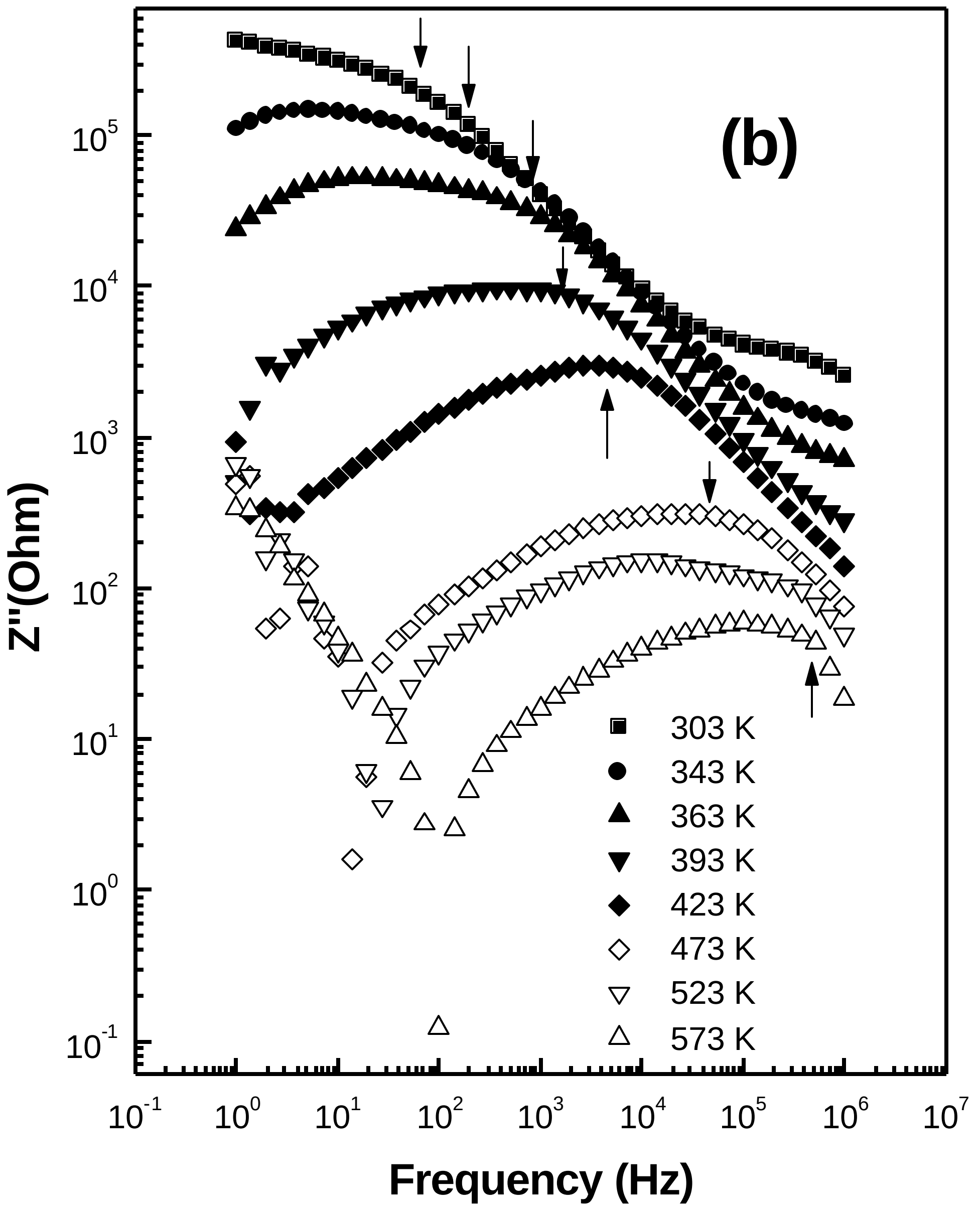

Fig. 5. Imaginary modulus (a) and imaginary impedance (b) spectra at different temperatures. The shift of the peak towards higher frequency with rise in temperature signifies dielectric relaxation.

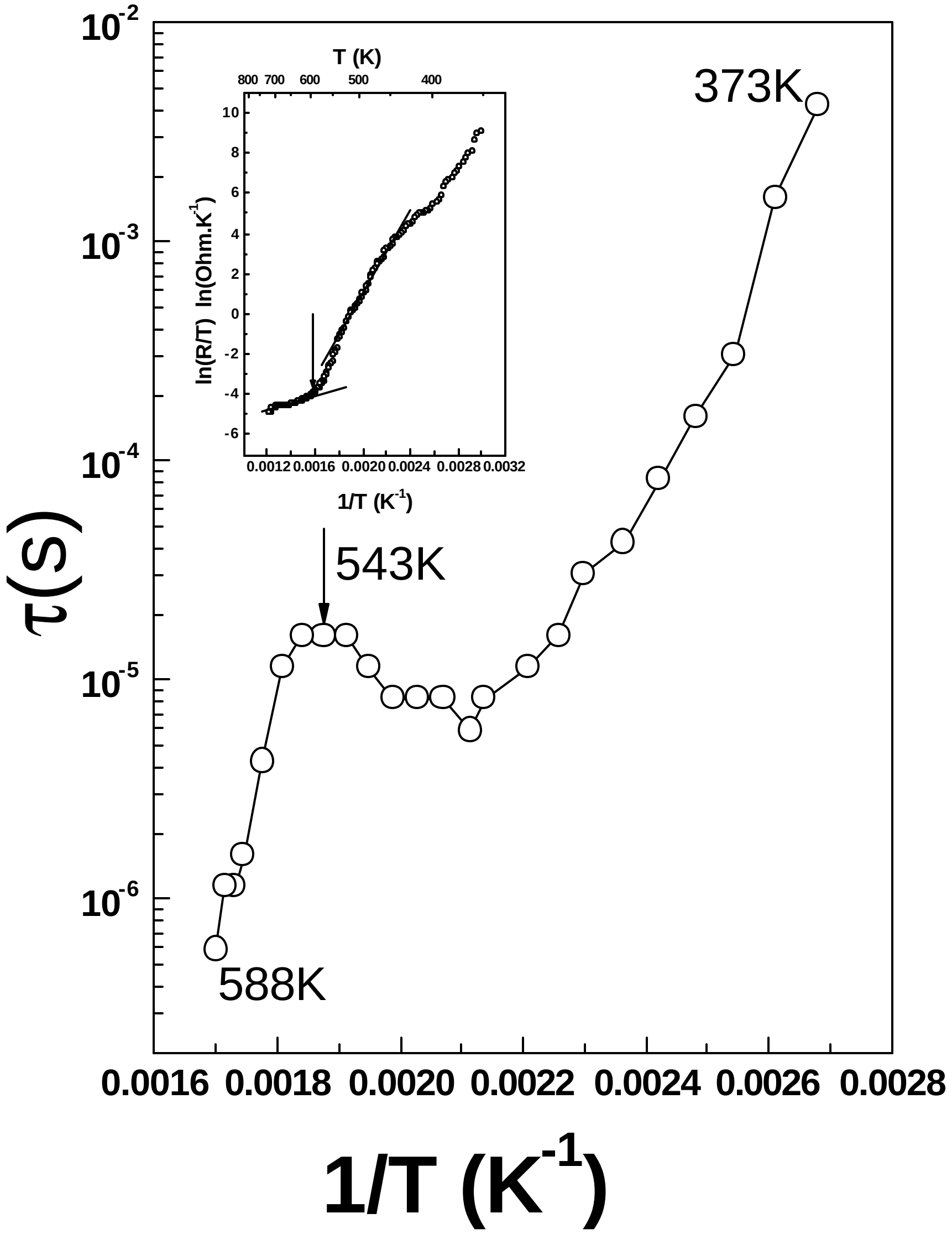

Fig. 6. The dielectric relaxation time (τ) vs. inverse temperature plot. In the inset the ln($\rho_{dc}$/T) vs. inverse temperature plot is shown.

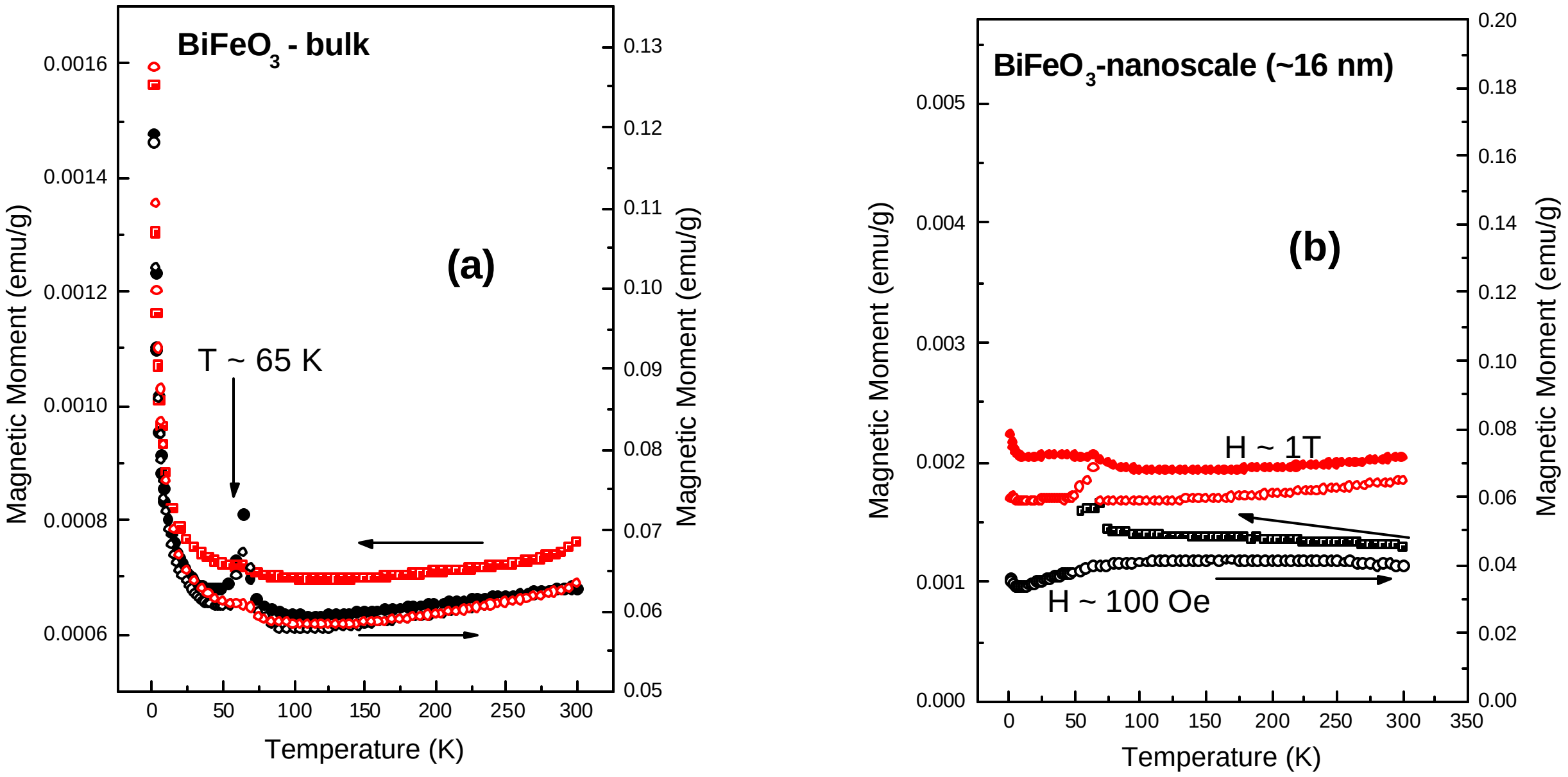

Fig. 7 (color online). The zero-field cooled and field cooled magnetization vs. temperature plots for the bulk and nanoscale $BiFeO_3$ under the applied fields – 100 Oe and 10000 Oe. The red colored plots are data for ~10000 Oe.

(a)

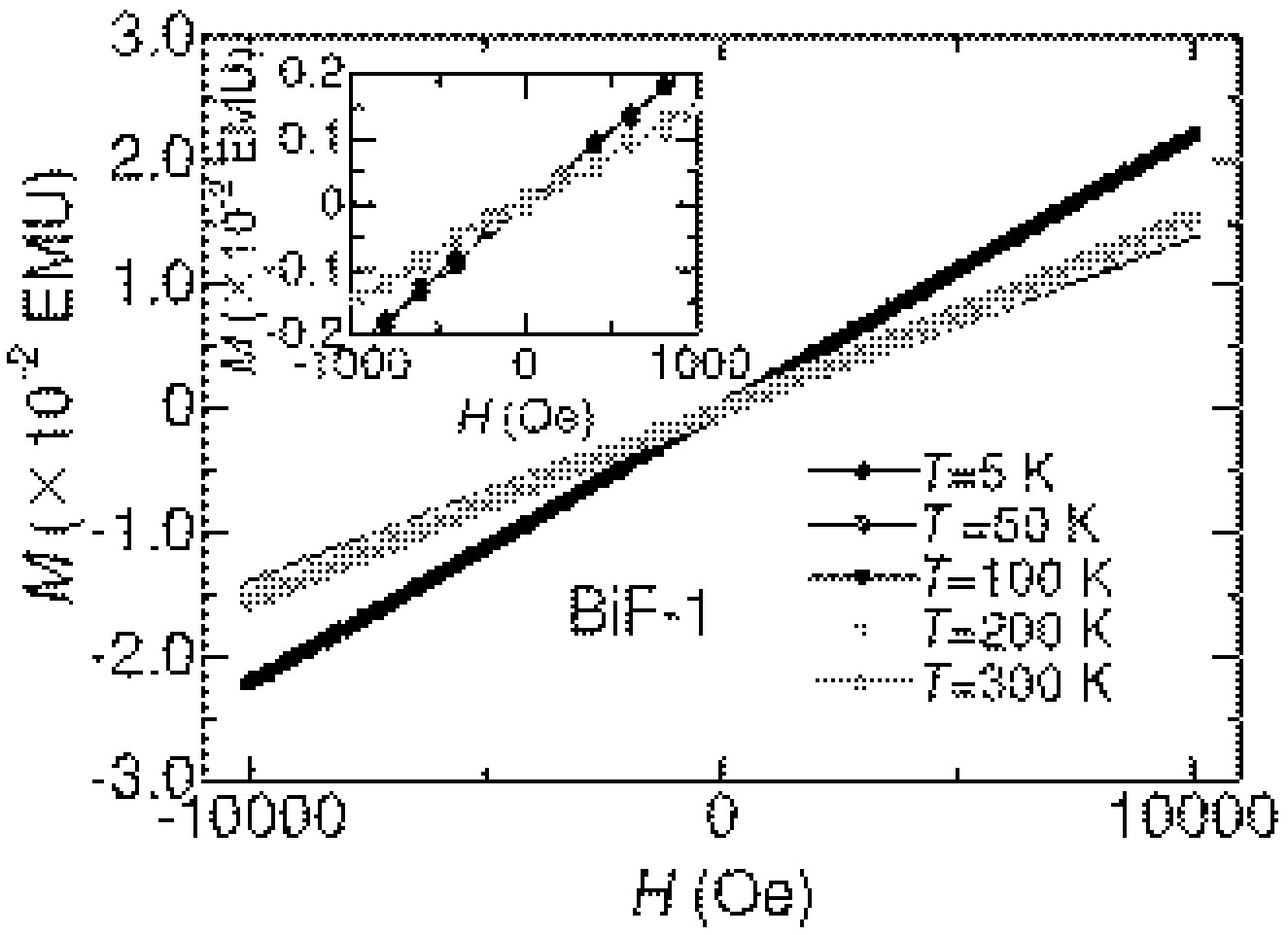

(b)

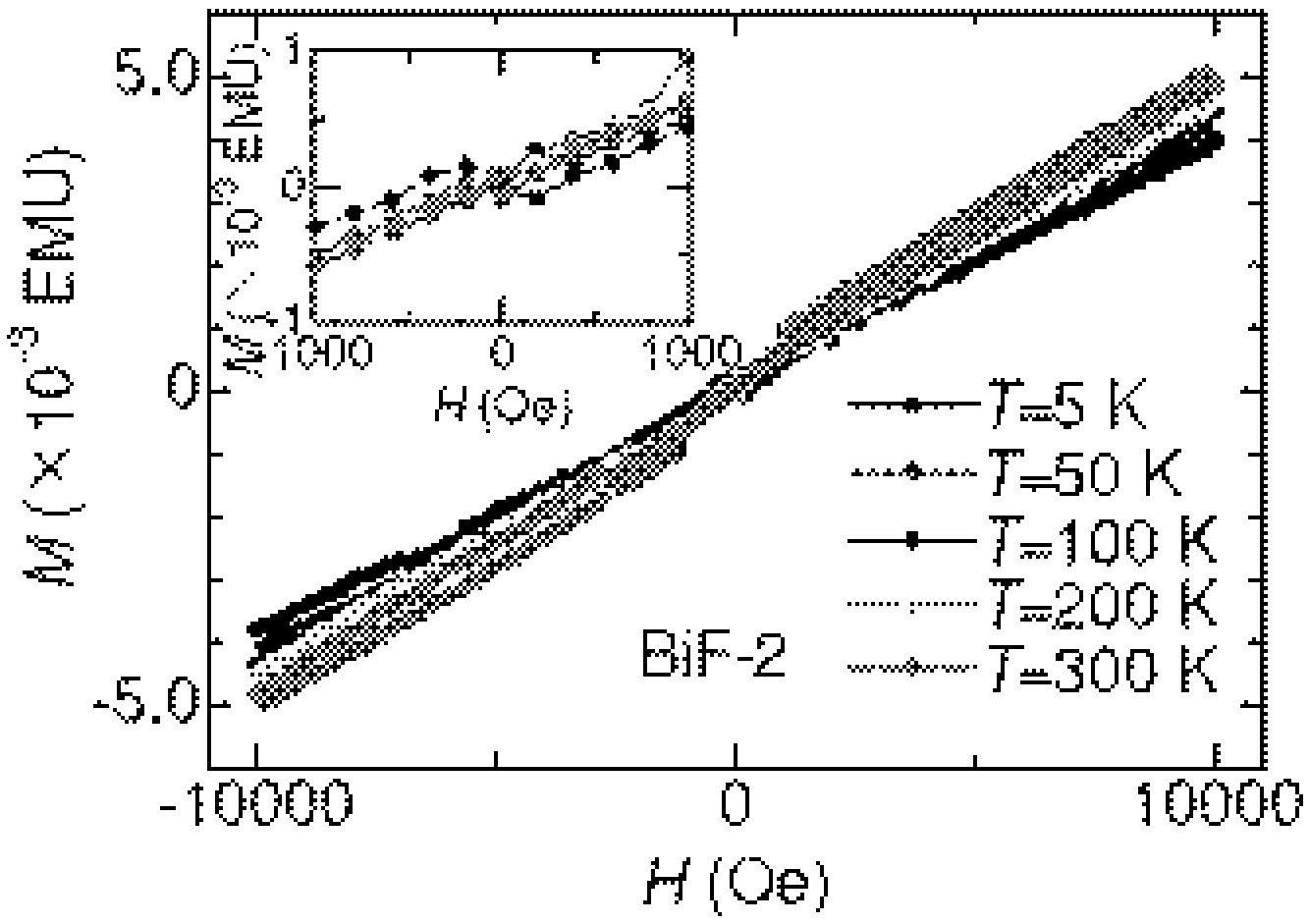

Fig. 8. The magnetic hysteresis plots for bulk (a) and nanoscale (b) $BiFeO_3$ at different temperatures.

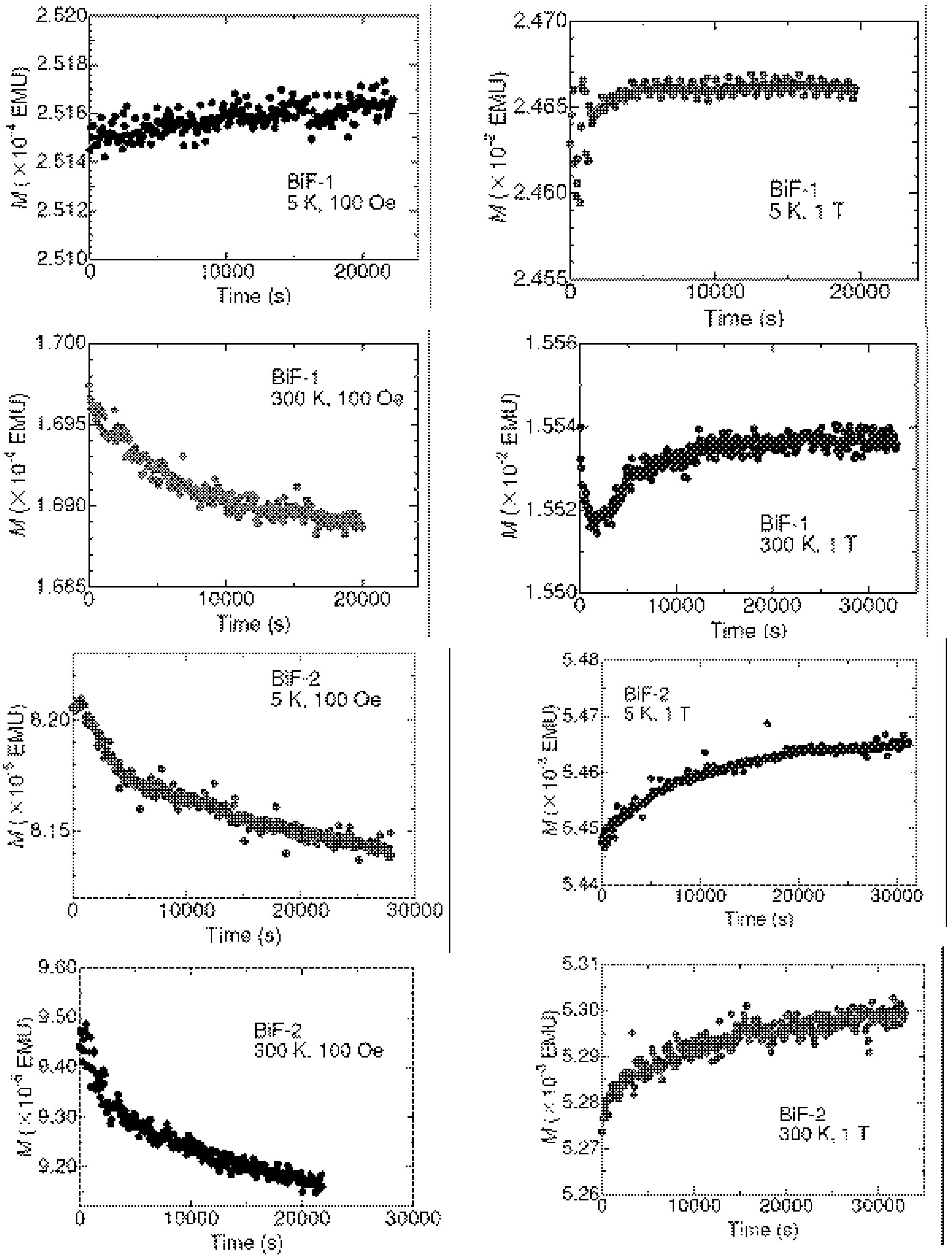

Fig. 9. The magnetic relaxation patterns for bulk and nanoscale $BiFeO_3$ under different fields and at different temperatures. BiF-1 and BiF-2 correspond to bulk and nanoscale samples, respectively.